\documentclass[prx,aps,amsmath,amssymb,twocolumn,superscriptaddress,10pt]{revtex4-2}
\usepackage{hyperref}
\hypersetup{
breaklinks=true,
colorlinks=true,
citecolor=blue,
linkcolor=magenta,
filecolor=blue,
urlcolor=blue
}

\usepackage{graphicx}
\usepackage{dcolumn}
\usepackage{bm}

\begin{document}

\preprint{APS/123-QED}

\title{Anomalous-diffusion synthesis of non-Gaussian reservoir anomalies for time-lapse seismic inversion
}

\author{Anderson Mateus de Sousa Nogueira}
\author{Paulo Vitor Ferreira}
\author{Katerine Rincon Perez}
\author{Jo\~ao M. de Ara\'ujo}
\affiliation{Department of Physics, Universidade Federal do Rio Grande do Norte (UFRN), Natal, RN, Brazil}

\author{Tiago Barros}
\author{Samuel Xavier-de-Souza}
\affiliation{Department of Computer Engineering and Automation, Universidade Federal do Rio Grande do Norte, Natal, Brazil}
\affiliation{Laboratory of Parallel Architectures for Signal Processing, Digital Metropolis Institute, Universidade Federal do Rio Grande do Norte, Natal, Brazil}

\author{Sérgio Luiz da Silva}
\affiliation{Laboratory of Parallel Architectures for Signal Processing, Digital Metropolis Institute, Universidade Federal do Rio Grande do Norte, Natal, Brazil}

\author{Gilberto Corso}
\thanks{Corresponding author}%
\email{gfcorso@gmail.com}

\affiliation{ Department of Biophysics and Pharmacology, Universidade Federal do Rio Grande do Norte, Natal, Brazil}



\date{\today}

\begin{abstract} 
We develop a physics-informed framework for learned time-lapse (4D) seismic inversion that estimates production-induced velocity changes associated with subsurface fluid migration. Existing learned inversion methods are commonly trained on seismic data generated from Gaussian diffusion models, which fail to capture the anomalous transport characteristic of heterogeneous subsurface reservoirs. To overcome this limitation, we introduce a data generator based on space-time fractional diffusion, in which the temporal and spatial fractional orders independently control memory effects and nonlocal transport, while elliptical anisotropy represents preferential flow. By sampling this parameter space, we generate a broad ensemble of physically consistent velocity perturbations spanning normal, subdiffusive, superdiffusive, and intermediate transport regimes. We then train a convolutional neural network to map time-lapse seismic residuals directly to velocity updates. Through extensive Monte Carlo validation, we show that the network accurately reconstructs production-induced anomalies across all transport regimes, with reconstruction errors governed primarily by anomaly amplitude and interface sharpness rather than by the diffusion regime itself. Our results demonstrate that embedding anomalous-transport physics into the training distribution substantially extends the applicability of amortized 4D seismic inversion, enabling robust recovery of heavy-tailed and anisotropic reservoir signatures beyond the Gaussian assumptions underlying existing approaches.
\end{abstract}

\maketitle


\section{Introduction}
\label{sec:intro}

Inverse problems are central to the physical sciences: from observations recorded outside a system, one seeks the internal parameters that produced them \cite{Razavy2020}. A growing body of work replaces the classical, per-instance optimization of such problems with amortized inference, in which a neural network is trained once to approximate the inverse map and thereafter returns a solution in a single forward pass \cite{karniadakis2021,lecun2015}. The appeal is acute wherever the forward model is expensive, and many inversions are required. It is also where the method is most fragile: a learned inverse operator is only as reliable as the distribution of examples it was trained on. In scientific settings, the labeled data are not collected but generated by simulation, and so the generative model of the physics is not a preprocessing detail---it is the effective prior that determines what the network can and cannot resolve. This paper examines that premise in the context of subsurface monitoring and argues, with a concrete construction, that grounding the generator in the correct diffusion physics unlocks generalization.

Reflection seismic, a geophysical method based on the emission of acoustic waves and the analysis of signals reflected at interfaces between rock layers with different acoustic impedances, constitutes the fundamental tool for subsurface imaging \cite{Tarantola2005,Yilmaz2001}. 4D seismic, or \textit{time-lapse} seismic, extends this technique by repeatedly acquiring 3D seismic surveys over the same field throughout the production life of a reservoir \cite{johnston2013}. Changes in the physicochemical properties of the reservoir induced by production and injection operations (water or $CO_2$) that modify fluid saturation and pressure, and consequently the acoustic impedance of the rock, generate so-called seismic anomalies between different seismic surveys \cite{nguyen2015}. The successful, accurate imaging of these anomalies is therefore directly correlated with the ability to monitor subsurface fluid dynamics, enabling the optimization of recovery strategies, the location of remaining hydrocarbon zones, and the validation of reservoir simulation models, thereby establishing itself as an important tool for production management and monitoring \cite{SAMBO2020103312,cruz_et_all_2021_4Dtupi}. The state-of-the-art deterministic approach, full-waveform inversion (FWI), fits a velocity model directly to the recorded wavefield by minimizing a least-squares misfit \cite{Sun_et_al_2003_PhysRevLett,virieux2009overview}. FWI is powerful but computationally demanding and notoriously non-convex: the misfit landscape is riddled with local minima (``cycle skipping'') \cite{daSilva_Kaniadakis_2022_PhysRevE.106.034113}, so convergence to the physical solution requires an accurate starting model and careful regularization \cite{lumley2001time}. These costs are multiplied in 4D, where the quantity of interest is a small differential update embedded in a much larger background \cite{daSilva_et_al_2024}.

Convolutional neural networks (CNNs) have emerged as an efficient alternative for seismic velocity-model building and inversion \cite{araya2018,yang2019,wu2020}. By learning a direct mapping from data to model, a CNN amortizes the inversion cost into a one-time training stage and exploits the translation-equivariant structure of wavefield data \cite{krizhevsky2012}. For 4D monitoring specifically, recent studies demonstrate that networks can map differential seismic data to impedance- or velocity-change models with competitive efficiency and useful robustness to noise \cite{rincon2024,eage_ferreira}. A common---and, we argue, limiting---feature of this prior work is the statistical model used to synthesize the training anomalies: production-induced changes are represented as smooth, Gaussian-shaped perturbations consistent with classical (Fickian) diffusion around wells. The network is then asked, at test time, to resolve changes drawn from the same idealized family.

High-fidelity 4D seismic imaging, crucial for monitoring fluid dynamics in reservoirs, faces the fundamental challenge of non-repeatability between surveys, where differences in data acquisition, environmental conditions, and noise attenuate the signal of true production anomalies \cite{johnston2013}. Advanced processing techniques, such as Full Waveform Inversion (FWI), seek to mitigate this problem by fitting velocity models directly to seismic data; however, the methodology is computationally expensive and requires sophisticated parameterization \cite{10.1190/geo2020-0577.1}. In contrast, modern machine learning approaches, particularly Convolutional Neural Networks (CNNs), have emerged as promising alternatives for 4D seismic inversion, offering a significant reduction in computational cost after an initial training stage. Recent work, such as that of \cite{rincon2024,eage_ferreira}, demonstrates substantial progress on this front, developing CNN architectures capable of mapping differential seismic data directly into impedance change models with greater efficiency and robustness to noise, thereby paving the way for more agile and accessible monitoring solutions.

The effectiveness of machine learning techniques, and CNNs in particular, strongly depends on a good training set; as a general rule, the generalization capacity of a neural network depends heavily on the representativeness of the training set \cite{scikit-learn,kong2019machine}. 4D seismic inversion is no different: 4D imaging is intrinsically dependent on the availability of robust and representative training models that encode reservoir properties in response to the seismic impulse. To build a comprehensive training database, it is imperative to generate a wide range of realistic seismic anomalies, simulating the variations that occur during production. This synthesis requires the use of sophisticated statistical techniques that go beyond the purely Gaussian models often employed to generate spatial variability in the medium. Indeed, the quality and diversity of synthetic anomalies are therefore fundamental to the generalizable performance of the neural network, enabling the CNN to learn to discriminate geological signal from noise and produce reliable subsurface images.

Fluid dynamics in hydrocarbon reservoirs can be described by different stochastic transport regimes, the nature of which is defined by the interaction between the porous medium's microstructure and the fluid molecules and the pressure gradients in the reservoir \cite{dake1983reservoir}. In the ideal case of a homogeneous medium with no dominant external gradients, molecular trajectories approximate Brownian motion, resulting in normal diffusion and a Gaussian distribution of displacements. However, the presence of a strong pressure gradient between injector and producer wells introduces a coherent advective component that shifts the mean of the distribution and breaks the symmetry of transport, opening the path to anomalous superdiffusion \cite{BOUCHAUD1990127}.

Conversely, in low-porosity and low-permeability zones, where rock-fluid interactions are intensified, particles undergo frequent retention and immobilization, transforming the random walk into a subdiffusive process in which the mean square displacement grows more slowly than in normal diffusion \cite{tiab2011petrophysics}.  A particularly complex scenario emerges in formations characterized by rocks with low effective porosity, such as certain dense carbonates or highly compacted sandstones \cite{dake1983reservoir}. In this context, reduced porosity is typically associated with very low permeability, significantly enhancing the physicochemical interactions between the rock matrix and the fluids \cite{berkowitz1995characterization,lenzi2023anomalous}. This phenomenon amplifies capillary and adsorption forces, effectively increasing the apparent fluid viscosity and introducing an additional resistance to flow \cite{civan2010effective}.

The transition between Brownian, superdiffusive, and subdiffusive regimes encapsulates the vast range of possible transport anomalies in real reservoirs. In the context of inverting reservoir anomaly properties using machine learning, it is important to consider all these possibilities in the training process so that a neural network can be built that is capable of handling these diverse scenarios and producing reliable responses.
In this context, we present a theoretical model using a generalized diffusion equation to simulate non-Gaussian anomalies that are used to train neural networks to produce time-lapse inversion. The rest of the paper is organized as follows. The methodology section is divided into three parts: the mathematical framework of the generalized diffusion, the machine learning strategy of inversion, and the seismic problem we address in the study. The results section shows both the non-Gaussian anomalies generated by generalized diffusion as well as the seismic inversion produced by the Machine Learning approach. In the last section we conclude the work and point out the limitations and the strong aspects of the paper.

\section{Methods}

The methodology of the paper is split into three parts. Initially, we address the challenges of modeling particle diffusion in a non-heterogeneous and complex reservoir---where anomalous transport behavior defies classical diffusion, and we employ fractional derivatives to capture the memory and non-local effects inherent to such porous media. Subsequently, a machine learning methodology based on Convolutional Neural Networks (CNNs) is developed to perform seismic inversion, leveraging the CNN's ability to extract spatial hierarchies and complex patterns from high-dimensional data. These methods are applied to the context of time-lapse seismic monitoring, where repeated seismic surveys are acquired over the same reservoir to detect dynamic changes in fluid saturation and pore pressure due to production or injection. In this framework, the seismic data consist of pre-stack volumes that record differences in acoustic velocities over time.

\subsection{Anomalous diffusion}
\label{sec:anomalies}

Classical diffusion describes the regime in which both the microscopic jump lengths and waiting times of transported particles possess finite mean and variance. In this context, the probability density $P(\mathbf{x},t)$ for an ensemble of particles satisfies Fick's second law,
\begin{equation}
\frac{\partial P(\mathbf{x},t)}{\partial t}
= K\,\Delta P(\mathbf{x},t),
\label{eq:fick}
\end{equation}
whose Green's function for this equation is Gaussian, and the mean-square displacement (MSD) increases linearly with time, $\langle |\mathbf{x}|^2\rangle \propto t$. This statistical behavior underlies the common modeling of reservoir anomalies as Gaussian distributions centered around wells. However, when the waiting-time or jump-length distributions exhibit heavy tails, as observed in trapping-rich or fracture-rich rock, the classical central-limit theorem does not apply. In such cases, the continuum limit of the underlying continuous-time random walk is governed by the space--time fractional diffusion equation \cite{montrollweiss1965,scher1975,metzlerklafter2000,metzler2004},
\begin{equation}
{}^{C}D_t^{\alpha_t}\,P(\mathbf{x},t)
= -K\,(-\Delta)^{\alpha_s/2}\,P(\mathbf{x},t),
\label{eq:fde}
\end{equation}
for $0<\alpha_t\le 1,\ 0<\alpha_s\le 2$. Here, $K>0$ is a generalized diffusivity, ${}^{C}D_t^{\alpha_t}$ represents the Caputo fractional derivative in time, and $(-\Delta)^{\alpha_s/2}$ denotes the Riesz fractional Laplacian in space. Both operators are nonlocal, acting in time and space respectively, and each reduces to its classical counterpart at the endpoints of its parameter range.

The Caputo derivative of order $0<\alpha_t\le 1$ encodes temporal memory: the
evolution at time $t$ depends on the entire history of $P$ through the
weakly singular kernel $
{}^{C}D_t^{\alpha_t} f(t)
= \frac{1}{\Gamma(1-\alpha_t)}
\int_0^{t} \frac{f'(\tau)}{(t-\tau)^{\alpha_t}}\,\mathrm{d}\tau$, where $\Gamma(\cdot)$ denotes the gamma function. When $\alpha_t=1$, the kernel reduces to a delta function and the operator becomes the ordinary time derivative, ${}^{C}D_t^{1} P = \partial_t P$, thereby recovering Eq.~\eqref{eq:fick}. In physical terms, $\alpha_t<1$ characterizes prolonged, heavy-tailed trapping events in confined pore spaces, corresponding to subdiffusion \cite{mainardi2010,gorenflo2014}.

The Riesz fractional Laplacian is most cleanly defined spectrally: it is the operator whose action multiplies the Fourier transform of a field by $|\mathbf{k}|^{\alpha_s}$,
\begin{equation}
\mathcal{F}\!\left\{(-\Delta)^{\alpha_s/2} f\right\}(\mathbf{k})
= |\mathbf{k}|^{\alpha_s}\,\widehat{f}(\mathbf{k}),
\quad
\widehat{f}(\mathbf{k}) \equiv \mathcal{F}\{f(\mathbf{x})\}.
\label{eq:riesz}
\end{equation}
When $\alpha_s=2$, this operator reduces to the classical Laplacian, $|\mathbf{k}|^{2}$. For $\alpha_s<2$, it becomes genuinely nonlocal and produces the heavy-tailed spatial jumps characteristic of Lévy flights, corresponding to superdiffusion \cite{klafter1987,metzlerklafter2000}. Therefore, $\alpha_t$ governs temporal memory (trapping), while $\alpha_s$ determines spatial nonlocality (long-range connectivity). Together, these parameters define a two-dimensional family of diffusion regimes.

Under periodic boundary conditions, Eq.~\eqref{eq:fde} is diagonal in the Fourier basis: each mode evolves independently, and its exact solution is the Mittag-Leffler propagator
\begin{equation}
\widehat{P}(\mathbf{k},t)
= E_{\alpha_t}\!\big(-K\,|\mathbf{k}|^{\alpha_s}\,t^{\alpha_t}\big)\,
  \widehat{P}(\mathbf{k},0),
\label{eq:mittag}
\end{equation}
where $E_{\alpha_t}(z)=\sum_{n=0}^{\infty} z^{n}/\Gamma(\alpha_t n+1)$ is the one-parameter Mittag-Leffler function. For $\alpha_t=1$ this collapses to the exponential $\exp(-K|\mathbf{k}|^{\alpha_s}t)$, recovering the classical propagator (Gaussian for $\alpha_s=2$), whereas for $\alpha_t<1$ its algebraic decay in $t$ encodes the memory of the Caputo operator. The anomaly field is reconstructed in physical space by the inverse Fourier transform of Eq.~\eqref{eq:mittag}.

The self-similar structure of Eq.~\eqref{eq:fde} is characterized by the scaling variable $|\mathbf{x}|/t^{\alpha_t/\alpha_s}$, indicating that the characteristic spreading length grows as $t^{H}$, where $H$ is the anomalous (Hurst) exponent,
\begin{equation}
H = \frac{\alpha_t}{\alpha_s}.
\label{eq:hurst}
\end{equation}
For $\alpha_s=2$, the second moment is finite and $H$ reduces to the conventional MSD exponent, $\langle|\mathbf{x}|^2\rangle\propto t^{\alpha_t}$, which yields subdiffusion when $\alpha_t<1$. In contrast, for $\alpha_s<2$, the spatial propagator exhibits algebraic tails and its variance diverges; in this regime, spreading is characterized by the self-similar scaling exponent~\eqref{eq:hurst} rather than by the MSD, and the bulk expands faster than linearly, indicating superdiffusion. The two exponents may also interact: $0<\alpha_t<1$ combined with $0<\alpha_s<2$ results in time-subordinated diffusion, where temporal trapping partially compensates for spatial flights, and the net spreading is determined by~\eqref{eq:hurst}. Table~\ref{tab:regimes} summarizes the canonical cases used to classify the ensemble.





\begin{table}[!htb]
\centering
\caption{Diffusion regimes spanned by the fractional-diffusion generator
and the parameter domains used to label the training ensemble. The
anomalous exponent is $H=\alpha_t/\alpha_s$
[Eq.~\eqref{eq:hurst}].}
\label{tab:regimes}
\begin{ruledtabular}
\begin{tabular}{lcc}
\textbf{Regime} & \textbf{Parameter domain} & \textbf{Spreading} \\
\hline
Normal diffusion           & $\alpha_t=1,\ \alpha_s=2$             & $H=1/2$ \\
Subdiffusion               & $0<\alpha_t<1,\ \alpha_s=2$           & $H<1/2$ \\
Superdiffusion    & $\alpha_t=1,\ 0<\alpha_s<2$           & $H>\tfrac12$ \\
Time-subordinated          & $0<\alpha_t<1,\ 0<\alpha_s<2$         & $H$ from Eq.~\eqref{eq:hurst} \\
\end{tabular}
\end{ruledtabular}
\end{table}

\begin{figure}[!htb]
\centering
\includegraphics[width=0.99\linewidth]{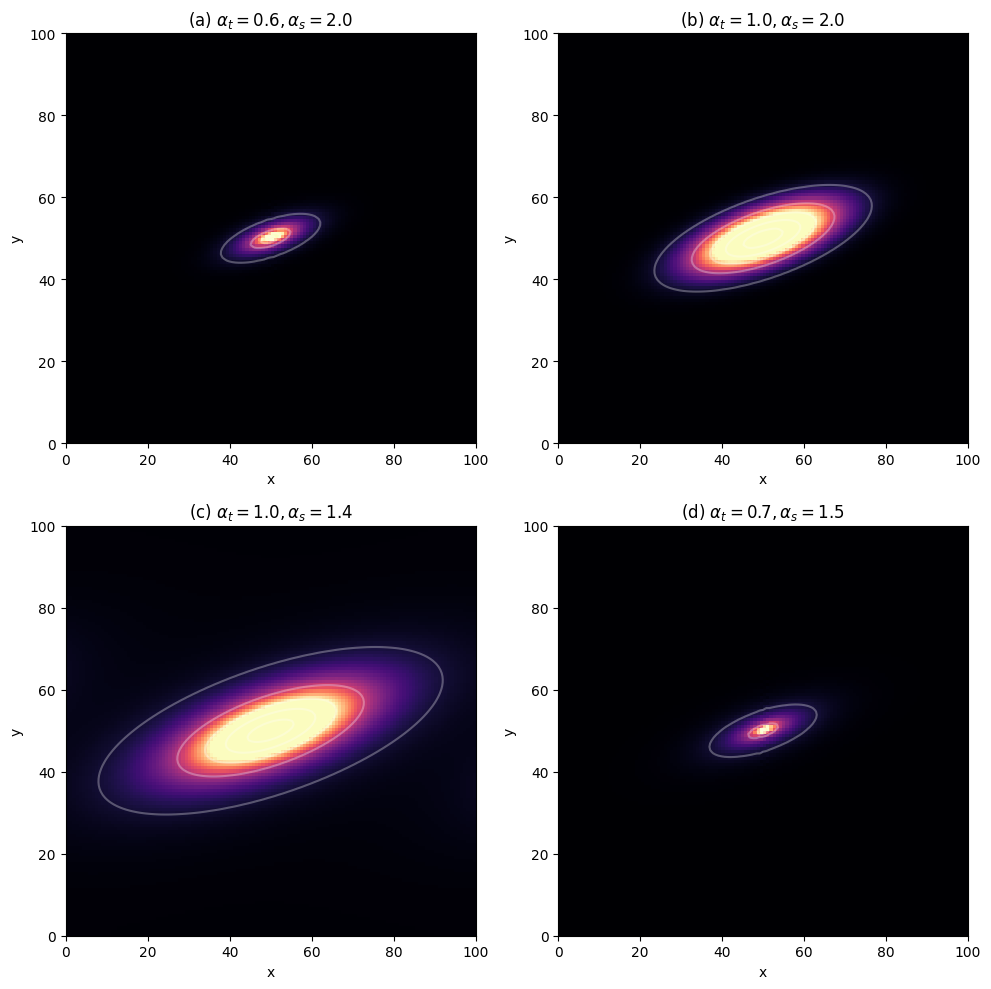}
\caption{Representative reservoir anomalies synthesized with the
fractional-diffusion generator [Eqs.~\eqref{eq:mittag} and
\eqref{eq:SFL_2d_eliptico}]; color scale logarithmic with a lower cutoff, and the
equiprobable surface outlined. (a) Subdiffusion, (b) normal diffusion,
(c) superdiffusion, and (d) time-subordinated superdiffusion. All cases
share the same elliptical anisotropy, isolating the effect of the
fractional exponents $(\alpha_t,\alpha_s)$ on plume morphology and tail
weight.}
\label{fig:anomalies}
\end{figure}

In practice, sedimentary media are directionally structured (bedding, aligned fractures), so isotropic spreading is the exception. In this regard, we introduce elliptical anisotropy by promoting the scalar diffusivity to a symmetric positive-definite tensor,
\begin{equation}
\mathbf{K}=
\begin{bmatrix} K_{xx} & K_{xy}\\ K_{yx} & K_{yy}\end{bmatrix},
\end{equation}
with $K_{xy}=K_{yx},\;\; K_{xx},K_{yy}>0,\;\; \det\mathbf{K}>0$. The spectral multiplier then becomes the anisotropic quadratic form
\begin{align}
    K|k|^{\alpha_s} &\xrightarrow[]{}  
     (\textbf{k}^\top \textbf{Kk})^{\alpha_s/2}
    \\&= \left( K_{xx} k_x^2 + 2K_{xy} k_x k_y + K_{yy} k_y^2 \right)^{\alpha_s / 2}
    \label{eq:SFL_2d_eliptico}
\end{align}
and the propagator~\eqref{eq:mittag} carries through with $K|\mathbf{k}|^{\alpha_s}$ replaced by $(\mathbf{k}^{\!\top}\mathbf{K}\mathbf{k})^{\alpha_s/2}$. The eigenvectors of $\mathbf{K}$ set the principal axes of the plume and the eigenvalue ratio its elongation, allowing the generator to mimic preferential flow directions. Figure~\ref{fig:anomalies} shows representative anomalies for the four regimes of Table~\ref{tab:regimes} under a common anisotropy, displayed on a logarithmic scale with the equiprobable contour outlined to make the heavy-tailed structure visible.

The field $P(\mathbf{x},t)$ describes the spatial pattern of production-induced change---the redistribution of the saturating or injected phase. To obtain the seismic observable we map this pattern to a perturbation
of the acoustic velocity. We adopt a linear rock-physics proxy in which the local velocity change is proportional to the normalized transport field,
\begin{equation}
\Delta V(\mathbf{x})
= \delta v_{\max}\,
  \frac{P(\mathbf{x},t)}{\max_{\mathbf{x}} P(\mathbf{x},t)} ,
\label{eq:proxy}
\end{equation}
with a prescribed peak amplitude $\delta v_{\max}$ sampled per realization. This proxy preserves the morphology and tail structure of the anomalous plume---the features that distinguish the diffusion regimes and that the network must learn to resolve---while abstracting the constitutive details of saturation-to-impedance coupling. A first-principles substitution (Gassmann fluid substitution \cite{gassmann1951,mavko2009}), which would tie $\delta v_{\max}$ and its spatial dependence to porosity, saturation, and pressure, is a natural refinement and is discussed in Sec.~\ref{sec:discussion}.

The generator exposes a small set of physically meaningful controls---the exponents $(\alpha_t,\alpha_s)$ that fix the regime, the tensor $\mathbf{K}$ that fixes the anisotropy and orientation, the snapshot time $t$ that fixes the extent, and the amplitude $\delta v_{\max}$. Sampling these controls produces a diverse ensemble of non-Gaussian anomalies $\{\Delta V_n\}$ spanning the four regimes of Table~\ref{tab:regimes}. We generate $N=5000$ such anomalies and confine each to the reservoir interval of the baseline model described in Sec.~\ref{sec:forward}, forming the targets for the learned inversion.

\subsection{Forward Problem and 4D Seismic Formalism}
\label{sec:forward}

The seismic inverse problem seeks the subsurface parameters that produced a set of recordings made at the surface or seabed. We write $\mathcal{M}\subseteq\mathbb{R}^{N_m}$ for the model space of those parameters and $\mathcal{D}\subseteq\mathbb{R}^{N_d}$ for the data space of recorded seismograms. Time-lapse monitoring is intrinsically differential: it compares two states of the \emph{same} reservoir and attributes their difference to production. We therefore introduce the 4D observable before the forward operator that generates it.

Let $\mathbf{m}_{\mathrm{base}}\in\mathcal{M}$ be the baseline model, acquired before or early in production, and $\mathbf{m}_{\mathrm{mon}}=\mathbf{m}_{\mathrm{base}}+\Delta\mathbf{m}$ the monitor model, acquired after a period of injection and withdrawal. The perturbation $\Delta\mathbf{m}$ is the velocity anomaly induced by the redistribution of fluids and pore pressure, synthesized by the anomalous-transport generator of Sec.~\ref{sec:anomalies}. Repeating the survey over each state yields data $\mathbf{d}_{\mathrm{base}}=\mathcal{F}(\mathbf{m}_{\mathrm{base}})$ and $\mathbf{d}_{\mathrm{mon}}=\mathcal{F}(\mathbf{m}_{\mathrm{mon}})$, where $\mathcal{F}:\mathcal{M}\to\mathcal{D}$ is the nonlinear forward operator defined below. The quantity that carries the monitoring signal is their difference, the time-lapse residual,
\begin{equation}
\Delta\mathbf{d}
= \mathbf{d}_{\mathrm{mon}} - \mathbf{d}_{\mathrm{base}}
= \mathcal{F}(\mathbf{m}_{\mathrm{base}}+\Delta\mathbf{m})
 -\mathcal{F}(\mathbf{m}_{\mathrm{base}}).
\label{eq:residual}
\end{equation}
The subtraction in Eq.~\eqref{eq:residual} is the crux of the 4D method: in the idealized limit of perfectly repeated acquisition, the response of the static, unchanged earth cancels exactly, leaving a residual sensitive only to the production-induced change $\Delta\mathbf{m}$. In practice the cancellation is imperfect---differences in source signature, receiver positioning, coupling, and ambient noise between vintages leave a \emph{non-repeatability} floor that competes with the true signal and constitutes the defining obstacle of field 4D monitoring \cite{johnston2013,lumley2001time}. The present study isolates the inverse-modeling question from this floor by working with repeated acquisition; we return to non-repeatability in Sec.~\ref{sec:discussion}.

The operator $\mathcal{F}$ is fixed by the physics of wave propagation. In a linear, isotropic, heterogeneous acoustic medium parametrized by the velocity $v(\mathbf{x}) \in \mathcal{M}$, the pressure wavefield $u(\mathbf{x},t)$ obeys the scalar wave equation
\begin{equation}
\mathcal{L}[v]\,u(\mathbf{x},t)
\equiv \left[\nabla^2 - \frac{1}{v^2(\mathbf{x})}\,
\frac{\partial^2}{\partial t^2}\right] u(\mathbf{x},t)
= s(\mathbf{x},t),
\label{eq:wave}
\end{equation}
where $\mathbf{x} \in \mathbb{R}^2$, $t$ denotes time, and $s(\mathbf{x}_s,t)$ is the source, a band-limited wavelet at the shot location $\mathbf{x} = \mathbf{x}_s$. The wave equation in~\eqref{eq:wave} is integrated with absorbing boundary conditions and sampled at the receiver array, so that the forward operator is the composition of the PDE solution map with a restriction operator $\mathcal{S}$ that records the wavefield at the $N_r$ receivers, $\mathbf{d}=\mathcal{F}(\mathbf{m})=\mathcal{S}\,u[m]$. The map is nonlinear in $m$ because the wavefield depends on the very medium through which it propagates.

However, recovering $\Delta\mathbf{m}$ from $\Delta\mathbf{d}$ is an ill-posed and an ill-conditioned problem. In this way, typically, $\Delta\mathbf{m}$ is obtained using methods such as FWI and Least-Squares Reverse-Time Migration (LSRTM), which are computationally very expensive. These are gradient-based methods and therefore require the determination of the Jacobian matrix. The Jacobian $\mathbf{J}$ carries a large effective null space: a finite array and a finite source bandwidth illuminate only a portion of the model wavenumber spectrum, so many distinct perturbations map to nearly identical data and the normal operator $\mathbf{J}^{\!\top}\mathbf{J}$ is severely ill conditioned. The standard deterministic remedy estimates $\Delta\mathbf{m}$ by minimizing a regularized least-squares functional,
\begin{equation}
\mathcal{J}(\Delta\mathbf{m})
= \tfrac12\,
\big\| \mathcal{F}(\mathbf{m}_{\mathrm{base}}+\Delta\mathbf{m})
 - \mathbf{d}_{\mathrm{mon}}\big\|_{\mathcal{D}}^{2}
 + \lambda\,\mathcal{R}(\Delta\mathbf{m}),
\label{eq:fwi}
\end{equation}
in which the data-misfit term enforces consistency with the observations, the stabilizer $\mathcal{R}$ (e.g.\ Tikhonov, $\mathcal{R}(\Delta\mathbf{m})=\|\mathbf{L}\Delta\mathbf{m}\|_2^2$) injects prior smoothness or sparseness, and $\lambda>0$ balances the two \cite{Tarantola2005,virieux2009overview}. This is the least-squares functional for FWI and LSRTM problems, for instance, and two features make its minimization costly. First, the operator $\mathcal{F}$ is strongly nonlinear and the misfit is non-convex: the oscillatory character of seismic data generates secondary minima (cycle skipping) that trap gradient-based methods unless the starting model already predicts arrivals to within half a period \cite{virieux2009overview,daSilva_Kaniadakis_2022_PhysRevE.106.034113}. Second, each gradient evaluation is expensive, so that each iteration requires at least two wave-equation solves per shot $s$. Incurred anew for every monitor survey, this cost---together with the
dependence on an accurate initial model---is precisely what motivates
replacing per-instance optimization with a single, amortized inverse
operator, learned once and evaluated in one forward pass
(Sec.~\ref{sec:ml}).


\subsection{The Seismic Model and Data Generation}

\begin{figure}[!htb]
    \centering
    \includegraphics[width = \linewidth]{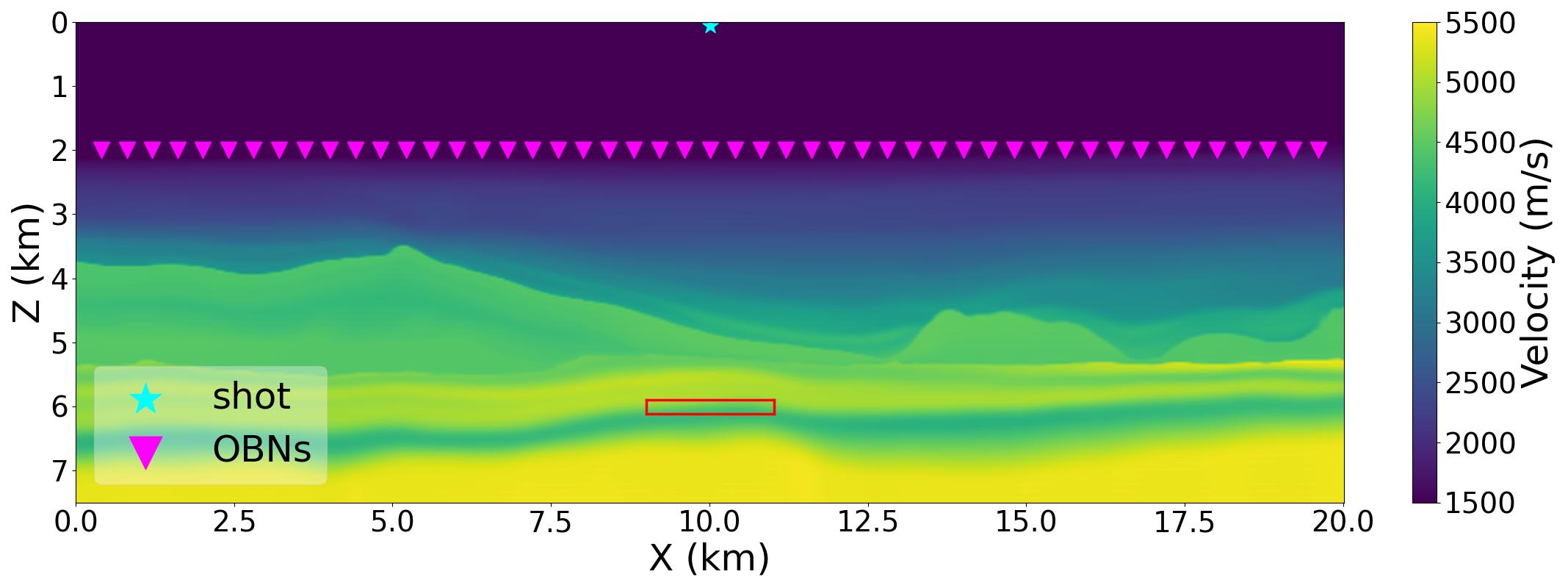}
    \caption{We employ a realistic 20 km wide, 7.5 km deep velocity model from a pre-salt oil field. The acquisition geometry consists of a single shot and 49 receivers. This velocity model is representative of typical pre-salt formations.}
    \label{fig:vel_model}
\end{figure}

To evaluate the proposed methodology, we consider a baseline velocity model representative of a Brazilian pre-salt formation \cite{cypriano2019,daSilva2021}; the model is 20 km wide and 7.5 km deep, with a reservoir target roughly 1.6 km thick and 4 km in lateral extent (Fig.~\ref{fig:vel_model}). The acquisition geometry consists of an Ocean Bottom Node (OBN) system with 49 receivers (nodes) distributed over a 20 km horizontal distance on the ocean floor, with a single-shot position located at the center of the velocity model on the sea surface. An overview of the baseline model and the acquisition setup is shown in Figure~\ref{fig:vel_model}. To isolate the variable under study---the statistics of the training anomalies---from confounding acquisition effects, we adopt a deliberately controlled OBN configuration: a single shot at the center of the model and $49$ receivers distributed over the 20 km aperture. 

The reservoir was positioned to represent a realistic geological scenario, with an approximate thickness of 1.6 km and a lateral extension of 4 km. To build the training dataset, this baseline model is used to generate 5000 synthetic monitor models considering four distinct non-Gaussian fractional diffusion scenarios. Each monitor model is obtained by introducing velocity anomalies ($\Delta V$), as described in Section~\ref{sec:anomalies}, within this reservoir region. 

For each synthetic monitor model, the corresponding seismic response is obtained via forward acoustic-wave simulations using the aforementioned acquisition geometry. The input data for the neural network consists of the normalized seismic differences ($\Delta S$) between the baseline and monitor datasets, gathered around the reservoir region.

\subsection{Deep Learning Inversion and CNN Architecture}
\label{sec:ml}

To address computational bottlenecks in iterative optimization for minimizing Eq.~\eqref{eq:fwi}, deep learning inversion reformulates the problem by introducing a parameterized inverse operator $\mathcal{G}_\theta : \mathcal{D} \rightarrow \mathcal{M}$, where $\theta$ denotes the trainable parameters of a deep neural network, including weights and biases. Here, $\Delta S \in \mathcal{D}$ represents the normalized seismic difference input, and $\Delta V \in \mathcal{M}$ denotes the predicted velocity update. The objective is to approximate the inverse mapping as follows:
\begin{equation}
    \Delta V \approx \mathcal{G}_\theta(\Delta S).
\end{equation}

\begin{figure}[!htb]
    \centering
    \includegraphics[width = \linewidth]{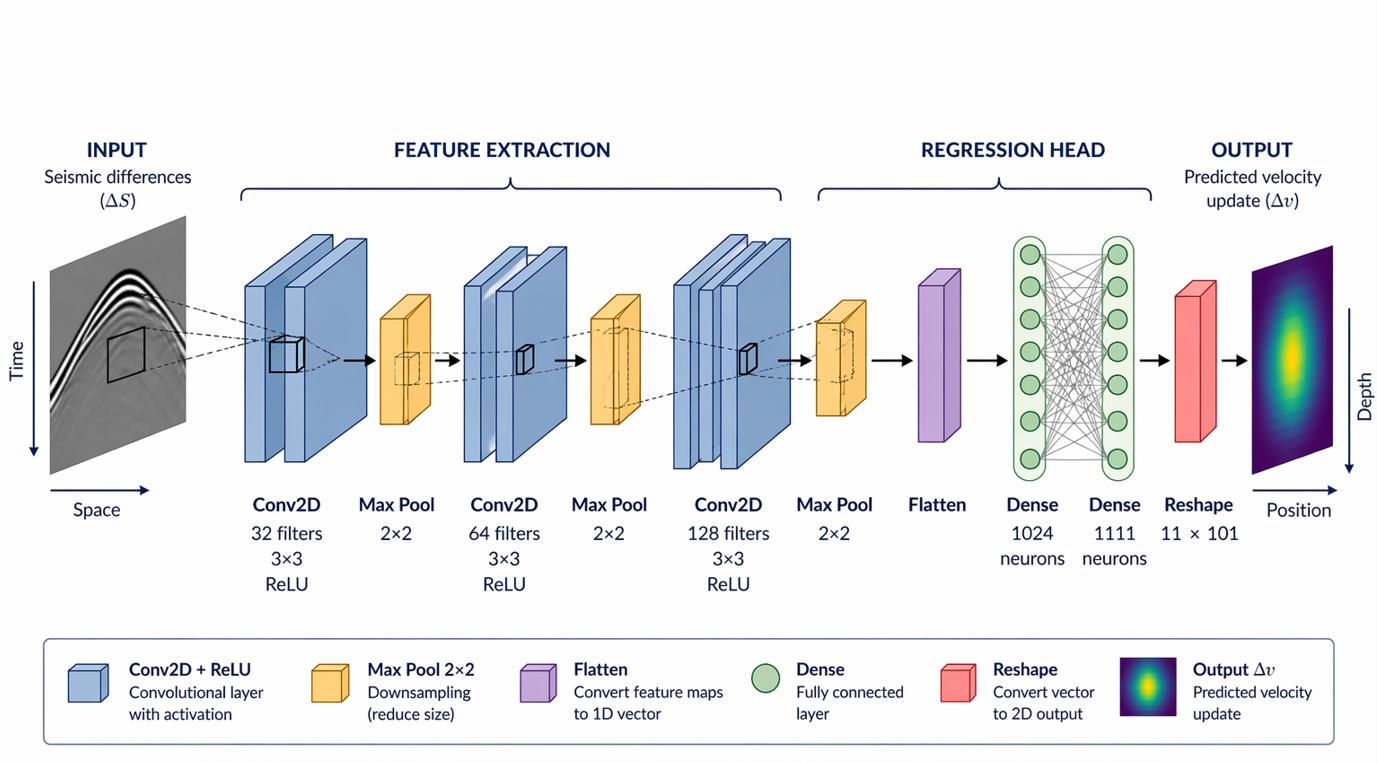}
    \caption{Proposed CNN architecture for velocity-update prediction from time-lapse seismic data. The network receives normalized seismic differences ($\Delta S$) between baseline and monitor datasets as input. Feature extraction is performed through successive Conv2D and max-pooling layers, followed by fully connected layers that map the extracted features into the predicted velocity update ($\Delta V$). The final reshape operation reconstructs the 2D velocity-update model within the reservoir region.}
    \label{fig:cnn_architecture}
\end{figure}

As illustrated in Figure~\ref{fig:cnn_architecture}, we employ a Convolutional Neural Network (CNN) architecture for its ability to exploit spatial correlations in seismic data through translation invariance. The network functions as a hybrid morphological mapper, comprising a feature-extraction module and a regression head.

The feature extraction phase includes sequential convolutional layers combined with spatial downsampling. The core operation in a discrete convolutional layer, which maps an input $I$ to a feature map $F$ using a kernel $K$, is defined as:
\begin{equation}
    F(i, j) = \sigma \left( \sum_{m} \sum_{n} I(i-m, j-n) K(m, n) + b \right),
\end{equation}
where $b$ is the bias term and $\sigma(\cdot)$ is a non-linear activation function, such as the Rectified Linear Unit (ReLU \cite{krizhevsky2012, lecun2015}). This stage progressively reduces the spatial dimensions of the residual data $\Delta S$ while increasing feature depth, extracting hierarchical, kinematic, and dynamic signatures from the wavefield.

After feature extraction, the resulting tensor is vectorized into a one-dimensional latent-space representation. This vector feeds the regression head, which consists of fully connected layers mapping the extracted features to the predicted velocity update. These layers perform global linear transformations followed by activation functions:
\begin{equation}
    \mathbf{a}^{(l)} = \sigma \left( \mathbf{W}^{(l)} \mathbf{a}^{(l-1)} + \mathbf{b}^{(l)} \right),
\end{equation}
where $\mathbf{W}^{(l)} \in \mathbb{R}^{N_l \times N_{l-1}}$ is the weight matrix for the $l$-th layer, $\mathbf{a}^{(l-1)} \in \mathbb{R}^{N_{l-1}}$ is the input activation vector from the previous layer, and $\mathbf{b}^{(l)} \in \mathbb{R}^{N_l}$ is the bias vector associated with the current layer $l$. The function $\sigma(\cdot)$ denotes the element-wise non-linear activation function. Consequently, $\mathbf{a}^{(l)}$ represents the output activation vector of the $l$-th layer, which will serve as the input for the subsequent layer.
The final fully connected layer contains $N_s$ neurons, corresponding to the spatial grid of the reservoir anomaly. The output is subsequently reshaped into the two-dimensional spatial domain to reconstruct $\Delta V$.

The network parameters $\theta$ are optimized iteratively via backpropagation. The dataset is randomly divided into training and test subsets containing 80\% and 20\% of the samples, respectively. Thirty independent training runs use Monte Carlo cross-validation to ensure statistical robustness. The loss function is the Mean Squared Error (MSE) over the entire spatial domain of the training batch:
\begin{equation}
    \mathcal{L}(\theta) = \frac{1}{N_{batch}} \sum_{k=1}^{N_{batch}} \| \Delta V_k^{\text{true}} - \mathcal{G}_\theta(\Delta S_k) \|_2^2.
    \label{eq:loss}
\end{equation}

While the network is trained holistically over the entire domain to capture the global mapping, the quantitative evaluation of predicted anomalies is conducted a posteriori. To ensure rigorous physical assessment and prevent the background medium (where $\Delta V = 0$) from artificially lowering the overall error, a localized binary mask is applied to the predictions. This restricts the MSE computation solely to the spatial support of the diffusion plume.

\section{Results}
\label{sec:results}


We assess the learned operator's ability to reconstruct the velocity update $\Delta V$ from the time-lapse residual $\Delta S$ across the four diffusion regimes in Table~\ref{tab:regimes}. We perform the reconstruction on the $11\times101$ reservoir grid ($N_s=1111$ output nodes) of the network (Fig.~\ref{fig:cnn_architecture}). We compute all statistics below on the held-out test predictions of the Monte Carlo cross-validation (MCCV), so no sample influences the model scoring it, and we read performance from the distribution over the $30$ folds rather than from any single split.

\begin{figure}[t]
\centering
\includegraphics[width=0.85\linewidth]{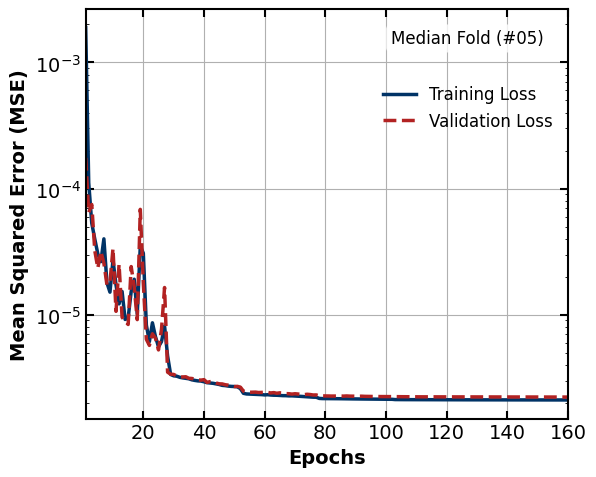}
\caption{Convergence history of the Convolutional Neural Network for the median-performing fold, selected from the 30 independent Monte Carlo cross-validation runs. The training (solid blue) and validation (dashed red) Mean Squared Error (MSE) losses are plotted on a logarithmic scale as a function of the training epochs.}
\label{fig:loss}
\end{figure}

Figure~\ref{fig:loss} presents the median-performing fold (fold~\#05): over $160$ epochs, both training and validation mean-square error [Eq.~\eqref{eq:loss}] decrease by approximately two and a half orders of magnitude, from $\sim\!10^{-3}$ to a plateau near $2\times10^{-6}$. After initial oscillations between epochs~$10$ and~$30$, the loss declines smoothly and converges by epoch~$60$. The validation curve closely follows the training curve throughout, with no late divergence, indicating that the operator generalizes across the anomaly family rather than memorizing specific cases. The low final loss demonstrates the network's capacity to fit the fractional-diffusion targets.

\begin{figure}[t]
\centering
\includegraphics[width=\linewidth]{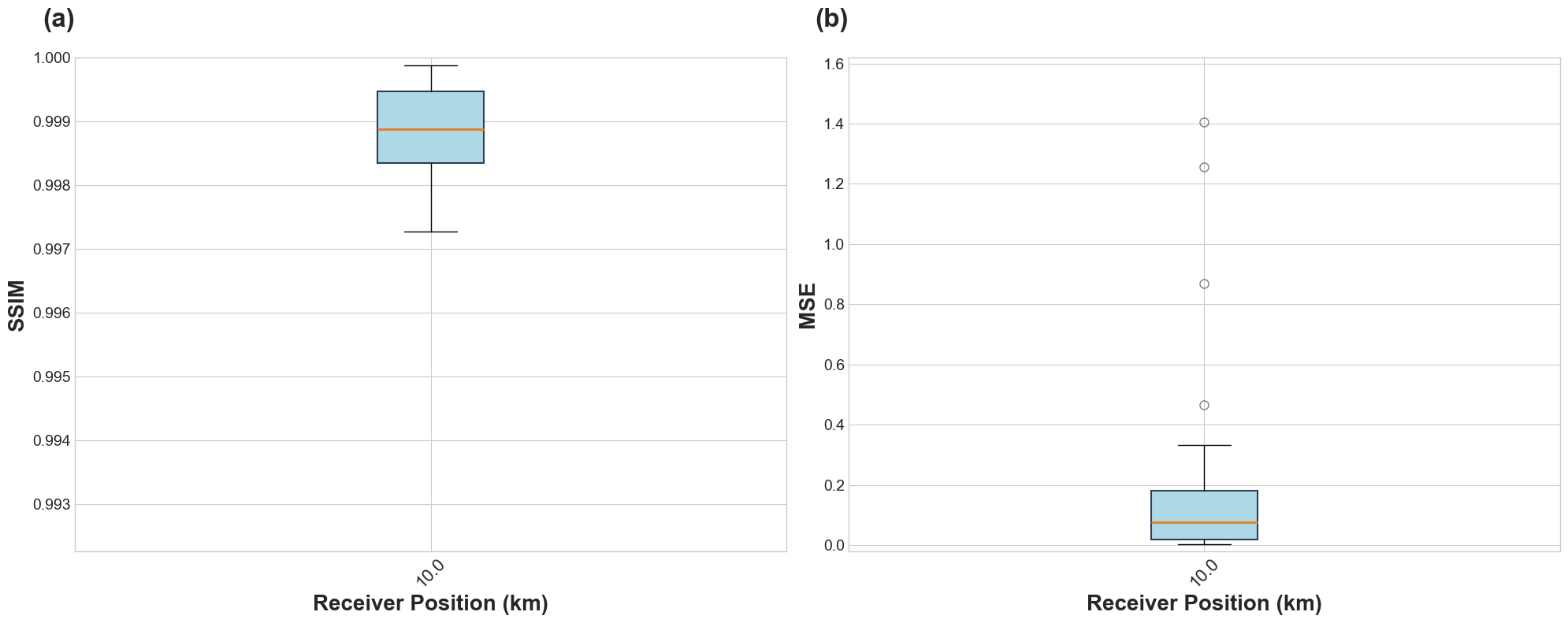}
\caption{Statistical evaluation of the CNN inversion performance over the 30-fold cross-validation at the 10.0 km receiver position. (a) Structural Similarity Index Measure (SSIM) distribution, highlighting the high spatial coherence and morphological fidelity of the predicted velocity anomalies. (b) Mean Squared Error (MSE) distribution, indicating the absolute pixel-wise error variance and the presence of localized outliers.}
\label{fig:boxplots}
\end{figure}

Expanding to the full ensemble, Figure~\ref{fig:boxplots} summarizes the two plume-restricted metrics across all $30$ folds. Structural similarity [Fig.~\ref{fig:boxplots}(a)] is consistently high: the median SSIM is $\approx\!0.999$, the interquartile range is $0.998$ to $0.9995$, and all folds exceed $\approx\!0.997$. Since SSIM measures local luminance, contrast, and structure \cite{wang2004}, these values indicate that predicted anomalies closely match the position, orientation, and shape of the true updates, regardless of the data split. The masked MSE [Fig.~\ref{fig:boxplots}(b)] is also low, with a median below $0.1$ and an interquartile range from $0.02$ to $0.18$, though a few outliers reach $\approx\!1.4$. These metrics are complementary: SSIM approaches unity because the morphology is almost always correct, while MSE highlights the few cases where amplitude is less accurately resolved. The narrow distributions confirm that performance reflects the method itself, not a favorable data split.

\begin{figure*}[t]
\centering
\includegraphics[width=0.86\linewidth]{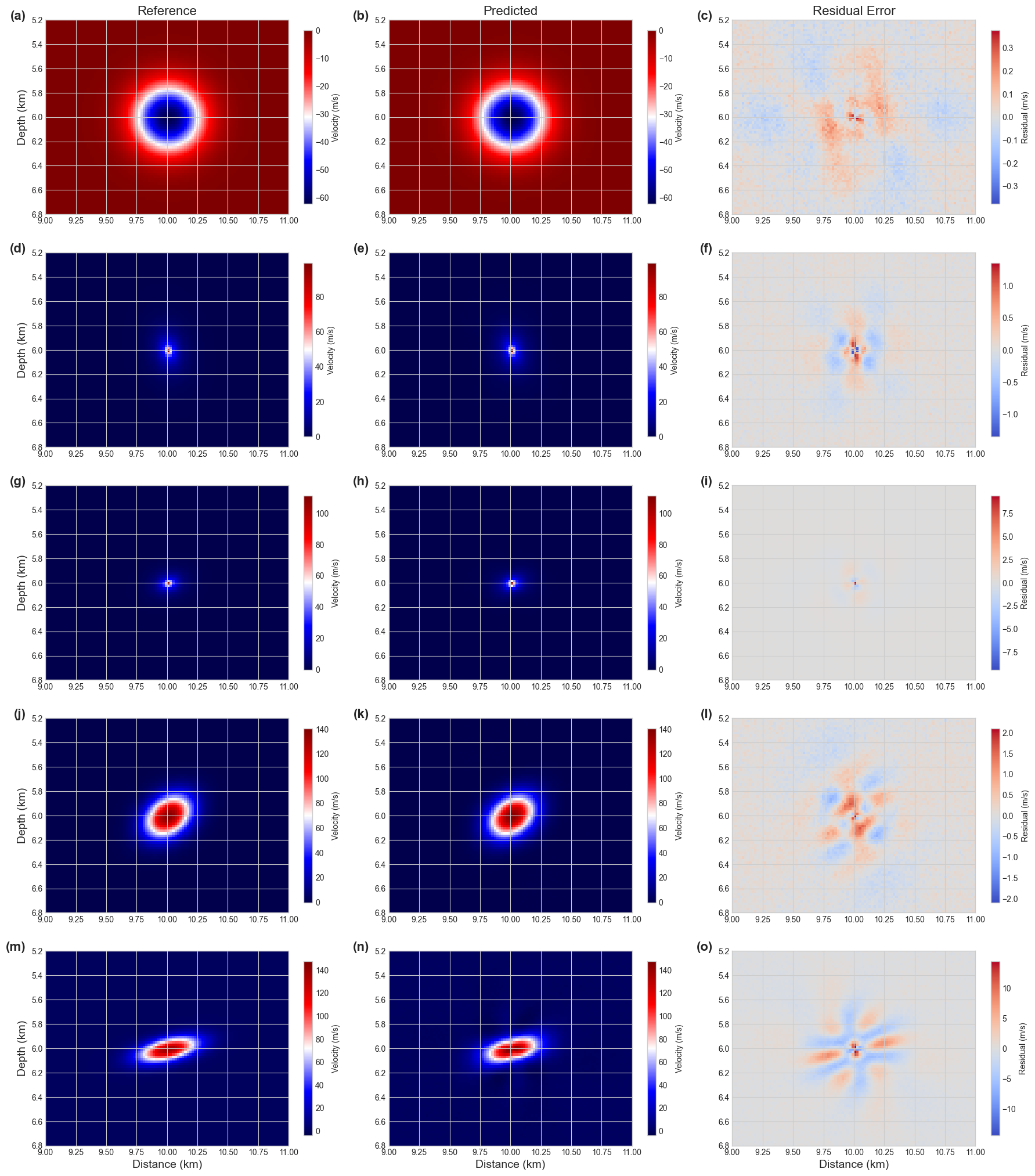}
\caption{Comprehensive visual analysis of the predictive performance across the Monte Carlo cross-validation folds for the velocity updates ($\Delta V$). To illustrate the full spectrum of the network's accuracy, the rows are organized according to the error distribution. From top to bottom: the first row (a-c) illustrates the best-case scenario, demonstrating excellent morphological reconstruction of the anomaly; the second row (d-f) represents the 25th percentile (first quartile) performance; the third row (g-i) presents the median-case scenario, reflecting the typical spatial coherence achieved by the model across the entire dataset; the fourth row (j-l) shows the 75th percentile; and the bottom row (m-o) displays the worst-case scenario, where the network struggled the most to resolve the complex heavy-tailed fractional diffusion. Columns from left to right correspond to the true reference anomaly, the CNN prediction, and the residual error (defined as the true minus the predicted velocity), respectively.}
\label{fig:spectrum}
\end{figure*}

To illustrate these statistics, Figure~\ref{fig:spectrum} presents held-out cases ranked by error, from best (top row) to worst (bottom row), showing the true anomaly, prediction, and residual (true minus predicted). The best case [Fig.~\ref{fig:spectrum}(a--c)] features a large, smooth, quasi-radial anomaly with moderate peak amplitude ($\sim\!60~\mathrm{m/s}$); the prediction is visually indistinguishable from the reference, and the residual is a low-amplitude speckle ($\pm0.3~\mathrm{m/s}$) without coherent structure. The first-quartile and median cases [rows (d--f) and (g--i)] are more compact and higher in amplitude ($\sim\!80$ and $\sim\!110~\mathrm{m/s}$), with residuals ($\pm1$ and $\pm7.5~\mathrm{m/s}$) localized at the anomaly core, indicating slight underfitting of sharp central gradients rather than misplacement. The third-quartile case [rows (j--l)] is a mildly elongated elliptical anomaly ($\sim\!140~\mathrm{m/s}$) with weak residual ringing at the rim. The worst case [Fig.~\ref{fig:spectrum}(m--o)] is a strongly elongated, obliquely oriented anomaly of similar amplitude, representing the anisotropic, heavy-tailed superdiffusive regime. Even in this case, the network recovers the core, position, and orientation; the largest residual ($\pm10~\mathrm{m/s}$) appears along the plume margins and tails, where steep, direction-dependent gradients are most challenging to fit.

Two trends are consistent across these cases. First, residuals remain confined to the anomaly core and margins, never appearing in the background. This supports the use of plume-restricted scoring and indicates that the operator has no systematic positional or orientational bias. Second, reconstruction difficulty depends more on peak amplitude and gradient sharpness than on the diffusion-regime label: large, smooth, low-amplitude plumes are reconstructed nearly perfectly, while compact or highly anisotropic high-amplitude plumes contribute to the upper tail of the MSE distribution in Fig.~\ref{fig:boxplots}(b). Because the masked MSE sums squared residuals over the plume support, a sharp but localized misfit, as in the median case, can yield a lower score than a milder error spread over a larger plume, as in the third-quartile case. This explains why peak residual amplitude does not always increase with error rank. The remaining errors are related to amplitude and edge resolution, not structure or location, which aligns with the near-unity SSIM.

\section{Final Remarks}
\label{sec:finalremarks}\label{sec:discussion}

The results from seismic inversion using convolutional neural networks (CNNs) trained on synthetic datasets have demonstrated good performance across both subdiffusive and superdiffusive transport regimes. The ML approach successfully captured the distinct signatures of anomalous diffusion: the heavy-tailed distributions and accelerated spreading associated with superdiffusive transport in high-permeability zones under strong pressure gradients, as well as the slower dispersion and particle retention patterns typical of subdiffusive regimes in low-porosity formations with intense rock-fluid interactions. By incorporating training examples that encode the full spectrum of fractional diffusion behaviors, from long-range connectivity effects to temporary particle immobilization in tight matrix sites, the neural networks learned to map seismic data directly from velocity change models. The results indicate that CNNs can effectively learn the complex, non-linear relationships between rock property distributions and fluid dynamics that govern production-induced seismic signatures.

The extreme lithological and structural diversity of hydrocarbon reservoirs gives rise to a broad spectrum of dynamical behaviors, extending beyond the scope of classical homogeneous models.  In heterogeneous media, the combination of advection, preferential channeling, complex connectivity, and permeability contrasts can produce effectively non-Fickian behaviors and, in some cases, superdiffusive spreading. 

The successful inversion of non-Gaussian anomalies represents a significant advancement beyond previously
published results limited to Gaussian-distributed reservoir changes \cite{rincon2024,eage_ferreira}.
While earlier work demonstrated the viability of ML for seismic inversion under idealized conditions assuming
normal diffusion and homogeneous property distributions, the current findings show that CNN architectures
can robustly handle the heavy-tailed, and spatially correlated patterns arising from genuine anomalous diffusion
processes. This capability is particularly significant for unconventional reservoirs where subdiffusive behavior
dominates due to nanoporous formations and complex fracture networks. To conclude, the adequate performance
across both subdiffusive and superdiffusive regimes confirms that it is possible to design neural
networks that properly model time-lapse seismic data to monitor reservoirs.

 \begin{acknowledgments}
The authors gratefully 
acknowledge NPAD/UFRN for the computational support,  FINEP Brazilian research agency  and CNPq Brazilian research agency for funding (grant no. 307907/2019-8, 314906/2023-1).
 \end{acknowledgments}



\nocite{}

\bibliography{referencias}

\end{document}